\documentclass[twocolumn,showpacs,amsmath,amssymb,superscriptaddress,prl]{revtex4}

\usepackage{graphicx}% Include figure files
\begin{document}

\title{Chaotic Free-Space Laser Communication over Turbulent Channel}

\author{N.F. Rulkov}
%\email{nrulkov@ucsd.edu}
\affiliation{Institute for Nonlinear Science, University of California, San Diego, La Jolla, CA 92093\\}

\author{M.A. Vorontsov}
\affiliation{Army Research Laboratory, Adelphi, Maryland 20783\\}

\author{L. Illing}
\affiliation{Institute for Nonlinear Science, University of California, San Diego, La Jolla, CA 92093\\}

\date{\today}

\begin{abstract}
The dynamics of errors caused by atmospheric turbulence in a
self-synchronizing chaos based communication system that stably
transmits information over a $\sim$5 km free-space laser link
is studied experimentally.
Binary information is transmitted using a chaotic
sequence of short-term pulses as carrier. The information signal
slightly shifts the chaotic time position of each pulse depending
on the information bit. We report the results of an experimental
analysis of the atmospheric turbulence in the channel and the
impact of turbulence on the Bit-Error-Rate (BER) performance of this
chaos based communication system.
\end{abstract}


\maketitle

Studies of chaos in nonlinear electrical circuits~\cite{ChaosComEC}
and lasers~\cite{ChaosComL} have shown that chaotic signals
generated in these systems can potentially be used as carriers for
information transmission. Thanks to the deterministic origin of
chaos, two coupled chaotic systems can self-synchronize reproducing
at the receiver~end the chaotic waveforms generated in the
transmitter~\cite{ChaosSynch}. This regime of self-synchronization
is a key element in the recovery of information encoded in the
received chaotic signal~\cite{ChCom}. Due to the variety and
complexity of the nonlinear dynamical issues involved, such chaos based
communication systems are of broad interest both for theoreticians
and experimentalists~\cite{spfi}.

All practical communication channels introduce signal distortions
that alter the chaotic waveform shape, as the result, the received
chaotic oscillations do not precisely represent the transmitter
oscillations. Channel noise, filtering, attenuation variability and
other distortions in the channel corrupt the chaotic carrier and
information signal. The presence of these channel distortions
significantly hamper the onset of identical synchronization of the
chaotic systems~\cite{SynchProb}. When signal distortions in the
channel exceed a certain level, self-synchronizing fails resulting
in failure of the communication link.

The enhanced sensitivity to chaotic signal waveform shape
distortions and the resulting problems with chaos synchronization
remain the major problems in the studies of chaos-based
communications systems. In order to overcome the problems of
channel distortions, a number of special chaotic communication
methods have been proposed~\cite{something}. At least in theory and
numerical simulations, it appears that the regime of identical
synchronization in these specially designed systems is
significantly less sensitive to channel noise and waveform
distortions caused by limited bandwidth of the
channel~\cite{something1}. However, to the best of our knowledge,
self-synchronizing chaos communication over a real-life highly
non-stationary channel has not been demonstrated so far.

In this letter we report the experimental study of a chaotic
self-synchronizing free-space laser communication in the presence
of severe communication signal distortions caused by atmospheric
turbulence. Chaotic pulse signals were used as the optical
communication carrier. Results demonstrate reliable
self-synchronization of two coupled chaotic systems for most of the
time and reveal the dynamical properties of errors bursts.
Synchronization failed only when deep signal fading occurred so
that the received power decreased to the photo-receiver noise
level.

\begin{figure}[h]
\includegraphics[width=.9\columnwidth]{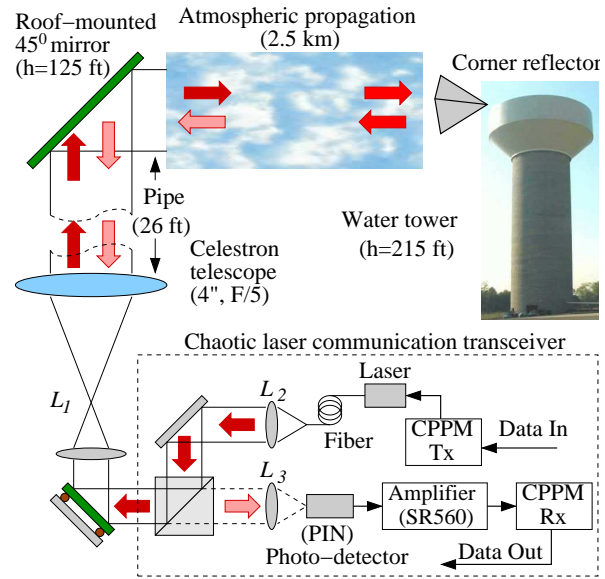}
\caption{Schematic for free-space laser communication system
based on chaotic pulse position modulation transceiver.}
\label{fig1}
\end{figure}

A schematic representation of the chaotic free-space laser
communication system used is shown in Fig.~\ref{fig1}. The
intensity-modulated 10 mW semiconductor laser beam ($\lambda
=690nm$) coupled to a single-mode fiber. Using a lens
relay system (lenses $L_{1}$ and $L_{2}$) and the transmitter
telescope (Celestron) the beam from the fiber was expanded to a
4'' diameter. The laser beam propagation path included a 26 ft long
vertical air-locked pipe connecting the optical table with a
$45^{\circ }$ mirror placed inside a shed on the roof of the
building, with subsequent propagation over an atmospheric path of
length $L\simeq 2.5$~km. At the end of the propagation path was a
$4"$ corner cube reflector placed on top of a water tower. After
reflection the laser beam propagated from the water tower back to a
communication receiver telescope in the roof-mounted shed. The
receiver system used the same Celestron telescope and lens relay
system (lenses $L_{1}$ and $L_{3}$) as did the transmitter system.
The total double-pass atmospheric laser beam propagation distance
was approximately $2L\simeq 5$~km long.

Double-pass wave propagation in a medium with random refractive
index fluctuations displays interesting statistical properties
known as backscatter enhancement.  Backscatter enhancement results
from correlations in the wavefront phase aberrations between the
outgoing and returned waves which have propagated through the same
refractive index inhomogeneities \cite{kravtsov93,andrews01}. The
variance of the received wave phase and intensity fluctuation
enhancement can exceed the corresponding value for a unidirectional
wave that propagates the distance $z=2L$ in an optically
inhomogeneous medium. Under conditions of strong intensity
fluctuations the backscatter enhancement factor can exceed a factor
of two \cite{saichev82}.

\begin{figure}[h]
\includegraphics[width=\columnwidth]{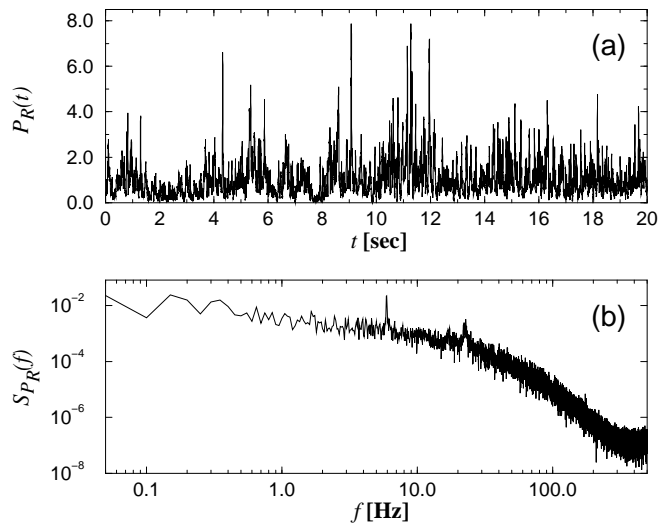}
\caption{Fluctuations of the received power $P(t)$ in the experiment
with non-modulated laser generating constant output intensity
(10mW). Normalized received power $P_R(t)=P(t)/\langle P(t)
\rangle$ measured at the photo-detector output~-~(a), and
corresponding averaged power spectrum of $P_R(t)$~-~(b) illustrate
the presence of strong laser beam intensity scintillations.}
\label{contturb}
\end{figure}

The received laser beam power was registered by the PIN
photo-detector (PDA55) placed in the lens $L_{3}$\ focal plane
(Fig.~\ref{fig1}). To evaluate the level of intensity
scintillations in the channel we examined the received signal from
a continuously running laser with a steady output intensity. An
example of the received signal fluctuations, measured by the PIN
photo-detector, amplified by the low-noise preamplifier (SR560 with
a gain of 20), and then acquired with sampling rate 1000
samples/sec, is presented in Fig.~\ref{contturb}a.

\begin{figure}[h]
\includegraphics[width=\columnwidth]{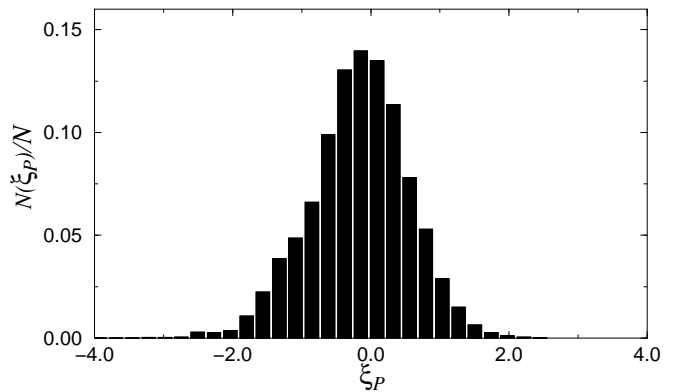}
\caption{Histogram of the probability distribution for the random
variable $\protect\xi _{P}=\ln (P/ \langle P \rangle)$ measured by
the PIN photo-detector. This histogram, $N(\xi_P)/N$, is computed
using $N=$ $10^{5}$ consecutive samples of the data $P(t)$, a 20
sec fragment of which is shown in Fig.~\ref{contturb}.}
\label{hist}
\end{figure}

The received signal standard deviation normalized by the mean value
is as high as 0.8-0.9, which is indicative of a {\it strong
scintillation regime}~\cite{andrews01}.            % <--- citation new resubmit2
 The corresponding ensemble-averaged received
signal power spectrum $S_{P_R}$ is shown in Fig.~\ref{contturb}b.
In atmospheric optics, laser beam intensity scintillations are
traditionally described in terms of the logarithm of the normalized
intensity $I$\ (for a point receiver) or received power $P$ (for
finite receiver telescope): $\xi _{I}=\ln (I)-\ln \langle I
\rangle$ or $\xi_{P}= \ln (P)-\ln \langle P \rangle$, where
$\langle I \rangle$ and $\langle P \rangle$ are ensemble (time)
averaged values \cite{andrews01,rytov89}. A histogram that
represents the distribution of the values of random variable
$\xi_{P}$ normalized by the total number of samples $N$, is shown
in Fig.~\ref{hist}.  Representing an approximation of the received
power probability distribution, the histogram in Fig.~\ref{hist}
closely matches the log-normal distribution expected from
theory~\cite{andrews01}.

In the experiment with chaotic transceiver the laser generated a
chaotic sequence of short-term ($\sim $1.0$\mu $s) on-off pulses of
intensity $U(t)=\sum_{j=0}^{\infty }w(t-t_{j})$. Here $w(t-t_{j})$
represents the waveform of an individual short-term rectangular
pulse generated at the moment of time
$t_{j}=t_{0}+\sum_{n=0}^{j}T_{n}$, where $T_{n}$ is the chaotic
time interval between the $n$th and the $(n-1)$th pulse. The laser
pulses were triggered by a TTL pulse signal from the chaotic
transceiver controller CPPM Tx (see Fig.~\ref{fig1}). The chaotic
sequence of the time intervals $\{T_{n}\}$ corresponds to
iterations of a chaotic process with the binary information signal
added to the chaotic signal. This method of chaos communication is
referred to as Chaotic Pulse Position Modulation
(CPPM)~\cite{CPPM}.
Since both chaos and information are in the timing of the pulses,
the particular intensity waveform of the generated light pulses is
of little consequence.

In the communication system discussed here the chaos is produced by
iterations of a one-dimensional tent map (see~\cite{CPPM} for
details of the hardware design). Time intervals in the generated
pulse sequence can be represented in the form of the following
iterative map:
\begin{equation}
T_{n}=F(T_{n-1})+d+mS_{n},  \label{map_enc}
\end{equation}%
where $F(\cdot )$ is a nonlinear function of the tent map and
$S_{n}$ is the binary information signal equal to either zero or
one. The parameter $m$ characterizes the modulation amplitude
whereas the parameter $d$ is a constant time delay needed for the
practical implementation. The nonlinear function $F(\cdot )$ and
parameters $d$ and $m$ were tuned to achieve a robust regime of the
map's chaotic behavior. The interpulse intervals $\{T_{n}\}$
fluctuated chaotically ranging
 from 10$\mu$sec to 25$\mu$sec and supported a $\sim$
60 kbit per sec bit-rate.

The distorted chaotic pulses $U'(t)$ received at the PIN photo
detector are applied to CPPM Rx (see Fig.~\ref{fig1}). Due to the
channel distortions and filtering in the PIN detector the received
pulses become of bell-shaped waveform. Each received pulse
triggered a timer circuit in CPPM Rx, when its amplitude exceeded a
certain threshold level, and the receiver acquired two consecutive
time intervals $T_{n-1}$ and $T_{n}$. The information signal was
recovered from the chaotic iterations $\{T_{n}\}$ using
formula~\cite{decoder}:
\begin{equation}
mS_{n}=[T_{n}-F(T_{n-1})-d].  \label{map_dec}
\end{equation}
Since the chaotic decoder map in the receiver is matched to the
encoder map in the corresponding transmitter, the time of the next
arriving pulse can be predicted (see Eq.(\ref{map_enc})). % <----- new resubmit2
To improve system performance by
reducing the probability of the channel noise falsely triggering
the decoder, the input of the synchronized receiver was blocked
until the moment of time when the next pulse was
expected~\cite{CPPM}.                   % <---------- citation is new (resubmit2)

\begin{figure}[h]
\includegraphics[width=\columnwidth]{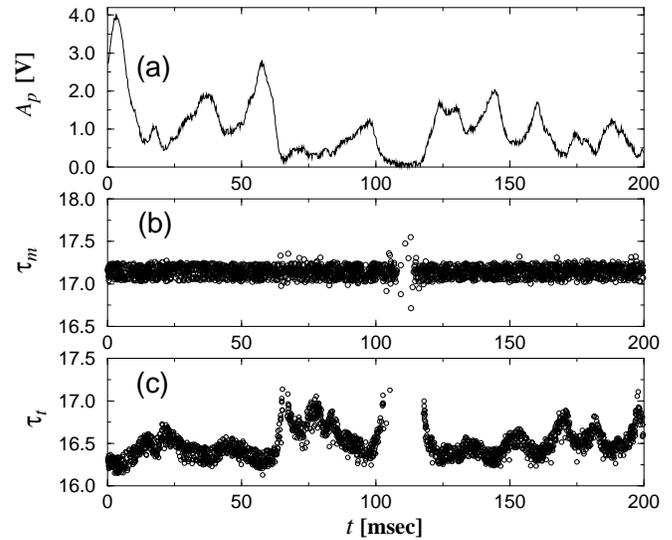}
\caption{Fluctuations of the CPPM pulses of light intensity
after traveling through atmospheric turbulence. Pulse amplitude
$A_p$ measured in volts at the output of amplifier~-~(a).
Propagation times $\tau_{m}$~-~(b) and $\tau_{t}$~-~(c) in
$\mu$sec. The pulse propagation times are computed from data
acquired simultaneously at the output of CPPM Tx and output of
Amplifier (SR560) at a sampling rate of 5$\times $10$^{6}$ samples
per sec.}
\label{atplot}
\end{figure}

Due to the effects of atmospheric turbulence the received pulses
were highly distorted. To illustrate, Fig.~\ref{atplot}a shows the
received pulse amplitude $A_{p}$ as a function of time. Despite the
severe pulse amplitude fluctuations clearly visible in
Fig.~\ref{atplot}a, the pulse propagation time $\tau _{m}$, which
is measured between the leading front of the TTL pulse applied to
the laser and the maximal point of the received pulse, varied only
within a 0.2$\mu$ sec time interval, see Fig.\ref{atplot}b. Small
fluctuations of the propagation time in the turbulent channel is a
potential for very good performance of CPPM communication method.
However, CPPM Rx is triggered when the leading front of the
received bell-shaped pulse waveform crosses the threshold level.
This level ($\sim$ 200 mV) was selected to minimize instances of
receiver controller triggering caused by noise, or by pulses
originating from local pulse reflections off nearby optical
surfaces. Therefore, the actual delay time $\tau_{t}$, measured
between the leading front of TTL pulses generated by CPPM Tx and
the moments of CPPM Rx triggering, depends on the amplitude of the
received pulses and fluctuates, see Fig.~\ref{atplot}c. Gaps in the
plots of $\tau_{m}$ and $\tau_{t}$ data occur due to the pulse
amplitude fading when the pulse amplitude falls to the
photo-receiver noise level and below threshold level, respectively.
Although $\tau_{t}$ changes with the amplitude variation these
changes remain less than the modulation amplitude $m\sim
1.5\mu$sec, see Eq.~(\ref{map_enc}). Slow and small variation in
the pulse propagation time is the key for CPPM controller
self-synchronization, and hence for stability of the entire
communication link.

\begin{figure}[h]
\includegraphics[width=\columnwidth]{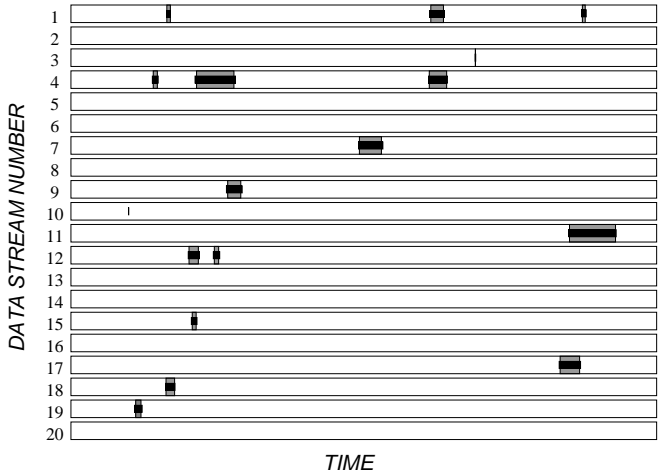}
\caption{Typical structure of errors shown in 20 consecutive
measured data streams each of length $\sim$170 msec transmitted at
$\sim$2~min intervals. Each strip presents 10000 bits which are
transmitted with the CPPM method. White intervals of the strips
mark blocks of data received without errors. Narrow black ribbons
in the middle of strips mark the blocks of the data received with
errors. The gray background shows the blocks of the dropped out
data caused by the loss of CPPM pulses due to fading instances.}
\label{fig4}
\end{figure}

The dynamics of errors caused by the atmospheric turbulence was
studied in the regime of real time transmission of binary
pseudo-random code data. An example of a map of the lost data in
such a transmission is presented in Fig.~\ref{fig4}. The total
Bit-Error-Rate measured in the experiment is $1.92\times10^{-2}$.
From the detailed analysis of the error structure we conclude that
main contributions to the BER are as follows. First, the loss of
bits carried by the pulses which did not trigger the CPPM receiver
due to the fading in the channel contributes
$\sim1.78\times10^{-2}$ to the BER ($\sim 92.7\%$). The fading
moments occur randomly during the communication and cause the drop
outs of blocks of data up to 1000 consecutive bits. The second
group of errors occurs in the relatively short time intervals right
before and after the failure of communication by fading. In these
time intervals the amplitude of the received pulses is still close
to the threshold and, as consequence, even small noise in the
channel can result in significant fluctuation of the interpulse
intervals (see Fig.~\ref{atplot}c). This effect contributes
$\sim1.4\times10^{-3}$ to the BER ($\sim 7.3\%$).
These two fading related error contributions would cause data loss
not only in this {\it chaos} based communication system but would
equally affect a similar type of {\it periodic} pulse position system.
The rest of the
errors which are not related to the complete failure of the channel
by the fading instances and can therefore be associated with the susceptibility of the {\it chaos} communication to the channel
distortions contributed to the BER only
$\sim5.5\times10^{-5}$.

This structure of
errors indicates that the CPPM communication method supports robust
communications over the turbulent channel except for the time
intervals when the channel fails due to fading. Thanks to the
self-synchronizing feature of this chaos communication method after
the total fading phase is over the CPPM receiver re-synchronizes
fast. In fact it needs to receive only two correct pulses to
establish the regime of chaos synchronization (see \cite{CPPM} for details).

The authors are grateful to L.S. Tsimring, H.D.I. Abarbanel, L.
Larson, and A.R. Volkovskii for helpful discussions. This work was
supported in part by U.S. Department of Energy (grant
DE-FG03-95ER14516), the U.S. Army Research Office (MURI grant
DAAG55-98-1-0269). The authors also thank \ J. Gowens and J.
Carrano for support in the development of the Atmospheric Laser
Optics Testbed (A\_LOT) at Adelphi, Maryland used in the
experiments.


\begin{thebibliography}{99}
\bibitem{decoder} In the CPPM Tx and CPPM Rx devices the consecutive
values of $T_n$ and $T_{n-1}$ are generated and stored in the form
of voltage signals. Eqs.~(\ref{map_enc}) and (\ref{map_dec}) are
implemented using an analog electrical circuit of the nonlinear
function $F(.)$ and a subtracting circuit, see Ref.~\cite{CPPM} for
details.
\end{thebibliography}
\end{document}